\documentclass[twocolumn,preprintnumbers,amsmath,amssymb,prl,superscriptaddress]{revtex4-1} 
\usepackage{amsmath} 
\usepackage[english]{babel} 
\usepackage{graphicx} 
\usepackage{listings} 
\usepackage{soul}
\usepackage{ulem} 
\usepackage{color}
\usepackage{lipsum}

\begin{document}

\title{Tuning the rheological behavior of colloidal gels through competing interactions}

\author{J. Ruiz-Franco}
\email[Corresponding author:]{ jose.manuel.ruiz.franco@roma1.infn.it}
\affiliation{Department of Physics, Sapienza University of Rome, Piazzale Aldo Moro 2, 00185 Roma, Italy}
\affiliation{CNR Institute of Complex Systems, Uos Sapienza, Piazzale Aldo Moro 2, 00185, Roma, Italy}

\author{F. Camerin}
\affiliation{CNR Institute of Complex Systems, Uos Sapienza, Piazzale Aldo Moro 2, 00185, Roma, Italy}
\affiliation{Department of Basic and Applied Sciences for Engineering, Sapienza University of Rome, via Antonio Scarpa 14, 00161 Roma, Italy}

\author{N. Gnan}
\affiliation{CNR Institute of Complex Systems, Uos Sapienza, Piazzale Aldo Moro 2, 00185, Roma, Italy}
\affiliation{Department of Physics, Sapienza University of Rome, Piazzale Aldo Moro 2, 00185 Roma, Italy}

\author{E. Zaccarelli}
\email[Corresponding author:]{ emanuela.zaccarelli@cnr.it}
\affiliation{CNR Institute of Complex Systems, Uos Sapienza, Piazzale Aldo Moro 2, 00185, Roma, Italy}
\affiliation{Department of Physics, Sapienza University of Rome, Piazzale Aldo Moro 2, 00185 Roma, Italy}

\date{\today}

\begin{abstract}
We study colloidal gels formed by competing electrostatic repulsion and short-range attraction by means of extensive numerical simulations under external shear. We show that, upon varying the repulsion strength, the gel structure and its viscoelastic properties can be largely tuned. In particular, the gel fractal dimension can be either increased or decreased with respect to mechanical equilibrium conditions. Unexpectedly, gels with stronger repulsion, despite being mechanically stiffer, are found to be less viscous with respect to purely attractive ones. We provide a microscopic explanation of these findings in terms of the influence of an underlying phase separation. Our results allow for the design of colloidal gels with desired structure and viscoelastic response by means of additional electrostatic interactions, easily controllable in experiments.
\end{abstract}

\maketitle
\section{I. INTRODUCTION}
Soft colloidal gels are ubiquitous in our everyday life and their fundamental study is crucial for advancements for biomedical, optical sensing or food-related industries~\cite{mezzenga2005understanding,dickinson2006colloid,gaponik2011colloidal,guvendiren2012shear,truby2016printing}. 
These low-density amorphous solids are characterized by weak interactions among the constituent colloids, allowing them to be easily manipulated through external stresses. Hence, to adapt them to our needs, it is important to gain an exquisite control on how to alter the spatial organization and the flow properties of the colloids that build up the solid.  

From the microscopic point of view, colloidal gels are characterized by the emergence of a percolating network~\cite{zaccarelli2007colloidal} into which particles are bonded through attractive interactions. These are usually controlled by means of depletion forces, which arise when nonadsorbing polymers are added to the suspension~\cite{lekkerkerker2011colloids}. In the absence of salt, such short-range attraction is usually complemented by a long-range repulsion of electrostatic origin. The presence of these two contributions gives rise to so-called competing interactions since attraction drives the aggregation while repulsion acts against it. Under these conditions, finite-size clusters can be stabilized at low packing fraction $\phi$~\cite{sciortino2005route}, originating equilibrium cluster phases~\cite{stradner2004equilibrium,sciortino2004equilibrium,zhang2012non} or arrested states known as Wigner glasses~\cite{toledano2009colloidal,klix2010structural}. However, on increasing $\phi$, such disconnected clusters merge into a percolating gel, but still mantain some indicative structural features of the underlying competing interactions, such as a prepeak in the structure factor~\cite{stradner2004equilibrium,cardinaux2007modeling} and a low fractal dimension~\cite{campbell2005dynamical,sciortino2005onedimensional}. It was shown that, in mechanical equilibrium, these gel properties can be tuned through simple changes in the competing interactions~\cite{valadez2013percolation,capellmann2016structure}. For instance, in the limit case where repulsion is suppressed, gel formation occurs through an arrested phase separation into a colloid-poor fluid and a colloid-rich glassy state~\cite{lu2008gelation}. This situation can be easily realized in experiments by screening the residual charges on the colloids. 

Detailed knowledge of the phase behavior in the presence of competing interactions has been fundamental for the study of biological systems such as proteins, including lysozyme~\cite{stradner2004equilibrium,godfrin2015short,riest2018short,bergman2019experimental} and antibody solutions~\cite{yearley2014observation,godfrin2016effect}, as well as for clay suspensions~\cite{ruzicka2010competing,angelini2014glass}. Likewise, the importance of soft colloidal gels for application purposes has positioned them as highly studied systems from the mechanical point of view~\cite{whitaker2019colloidal,tsurusawa2019direct}. Nowadays, we have at our disposal a great deal of information regarding the rheological behavior of gels obtained by simple attractive interactions, encompassing (among others) the presence of two yielding points under steady shear~\cite{koumakis2011two,chan2012two,laurati2011nonlinear}, accumulation of residual stresses~\cite{moghimi2017colloidal} and non-Newtonian behavior~\cite{osuji2008shear,divoux2013rheological}. Instead, investigations of the rheological behavior of systems with competing interactions have been surprisingly scarce~\cite{imperio2008rheology,krishna2012probing,kadulkar2019influence,ruiz2019rheological,kadulkar2019influence}. 

To fill this gap, we investigate -- by means of extensive numerical simulations -- the behavior of colloidal gels induced by competing interactions undergoing a start-up shear test. In particular, we tune the contribution of the long-range repulsive term and find important differences in the evolution of the stress stored in the system. From this analysis, we are able to show that the complex interplay between short-range attraction and long-range repulsion allows the modification of the resulting gel properties, including its viscosity and the final structure in the steady-state regime. In particular, our findings demonstrate that the gel fractal dimension can be manipulated to a great extent in the presence of shear and we also provide arguments to predict some aspects of the large strain behavior from properties calculated at yielding. Finally, and most strikingly, we find that the addition of the long-range repulsive term makes the system stiffer, but less viscous with respect to the purely attractive case. These counterintuitive findings can be easily rationalized in terms of the proximity of the system to an underlying phase separation.  Our results can be readily verified in experiments through of a simple variation of the effective charge of the interacting particles. 

\section{II. NUMERICAL METHODS}
\subsection{A. Simulation models}

We simulate a $50:50$ binary mixture system of spheres with unit mass $m$ of diameters $d_{A}=0.9d$ and $d_{B}=1.1d$, interacting via a potential $V\left(r\right)$ which is the sum of a short-range attraction and a long-range repulsion~\cite{sciortino2004equilibrium,sciortino2005onedimensional},

\begin{equation}
\label{eq:SR}
V(r)=4\epsilon\left[\left(\frac{d}{r}\right)^{2\alpha}-\left(\frac{d}{r}\right)^{\alpha}\right]+A \frac{e^{-r/\xi_{D}}}{r/\xi_{D}}\,,
\end{equation}

\noindent with $r$ being the distance between two particles and $d=\left(d_{A}+d_{B}\right)/2$. Here, the short-range attraction is modeled by a generalized Lennard-Jones potential~\cite{vliegenthart1999strong}, being $\epsilon$ the potential depth. On the other hand, the long-range repulsion is described by a Yukawa contribution, where $\xi_{D}$ is the Debye screening length, and $A$ is proportional to the effective charge of the colloidal particles. The two contributions in $V\left(r\right)$ aim at representing, respectively, the depletion attraction and a screened electrostatic repulsion often observed in colloidal systems. Following Ref. \cite{sciortino2005onedimensional}, we fix $\alpha=18$ and $\xi_{D}=0.5d$, whereas we vary $A$  in order to manipulate the proximity to phase separation of our colloidal gels (see Fig. 1(a) to observe the influence of $A$ on $V\left(r\right)$). In particular, we study the rheological behaviour of the system for $A\in\left[0, 2, 3, 4, 6, 8\right]\epsilon$. In our simulations, length, mass and energy are measured in units of $d$, $m$, and $\epsilon$, respectively. Thus, time is measured in units of $\tau=\sqrt{md^{2}/{\epsilon}}$. In addition, we set $k_{B}=1$ and use a cutoff for the interactions at $r_{c}=8\xi_{D}$. 

Our simulations focus on a packing fraction $\phi=\frac{\pi}{6} \langle d^{3} \rangle \frac{N}{V}=0.16$, where $V$ is the volume of the cubic simulation box. To reduce size effects, all simulations are carried out for systems composed of $N=50000$ particles. We start from an equilibrated system at $T=1.0$ to remove defects due to the random generation of our initial configurations. Then, we quench the system down to $T=0.1$, which we then maintain fixed throughout. At this low $T$, the system becomes arrested either in a spinodal gel (for low values of $A$) or in an equilibrium gel (for larger $A$). Next, we perform a start-up shear test by applying a steady shear flow $\dot{\gamma}$ onto the gel imposing the Lees-Edwards boundary conditions~\cite{lees1972computer}. Here, $\dot{\gamma}$ is given in units of $\tau^{-1}$, and hence, the strain $\gamma={\dot{\gamma}\tau}$ is dimensionless. We consider the gradient velocity to be in the $\hat{y}$ direction while the shear velocity is in the $\hat{x}$ direction. In this way, the shear rate is defined as $\dot{\gamma}\equiv v_{x}/y$. 

\subsection{B. Equations of motion}

All simulations are performed in the $NVT$ ensemble using the Langevin equation, where the total force on the $i-$th particle is defined as

\begin{equation}
\label{eq:Langevin}
\mathbf{F}_{i}=\mathbf{F}^{C}_{i}+\mathbf{F}^{D}_{i}+\mathbf{F}^{R}_{i}\,.
\end{equation}

\noindent with $\mathbf{F}^{C}_{i}$ being the conservative force computed from the interaction potential defined in Eq.~(\ref{eq:SR}). Affine deformations are considered to act on the dissipative forces and specifically on the peculiar velocity~\cite{shang2017assessing,ruiz2018effect}, i.e., $\mathbf{F}^{D}_{i}=-\xi v'_{i,x}=-6\pi\eta_{s} d\left[v_{i,x}-u_{x}\left(y\right)\right]$. Here $\xi$ is the friction coefficient, $\eta_{s}$ is the viscosity of the implicit solvent, and $u_x(y)=\dot{\gamma} y$ is the stream velocity. In particular, we fix $\xi=10^{2}$. Finally, random forces $\mathbf{F}^{R}_{i}$ are defined with zero mean, $\left\langle F_{i}^{R}\right\rangle=0$, and to be $\delta$-correlated, $\left\langle F_{i}^{R}\left(t\right)\cdot F_{i}^{R}\left(t'\right) \right\rangle=2k_{B}T\xi\delta\left(t-t'\right)$. The equations of motion are integrated using a time-step $dt=0.002$. Simulations were performed with LAMMPS~\cite{lammps}.

\subsection{C. Observables}

{\it Shear stress tensor.} In the presence of the steady shear flow, we calculate the internal shear stress tensor $\sigma_{xy}$ using the Irving-Kirkwood expression~\cite{irving1950statistical}:

\begin{equation} 
\label{eq:Stress_Tensor}
\sigma_{xy} = \frac{1}{V}\left\langle\sum_{i}[m_{i}v'_{i,x}v_{i,y}+\sum_{j>i}r_{ij,x}F_{ij,y}]\right\rangle\,,
\end{equation}

\noindent where $r_{ij}$ and $F_{ij}$ are, respectively, the distance and the force between particles $i,j$ and the brackets $\langle \ldots \rangle$ represent the ensemble average. The first term captures the kinetic contribution from the shear flow, whereas the second one describes the configurational distribution of the first neighbours with respect to the $i-$th particle.

{\it Demixing parameter.} To quantify the phase separation of the colloidal gels, we divide the simulation box into $l^{3}$ cells of equal length with $l$ the number of cells in one dimension. Thus, we compute the demixing parameter $\Psi_{l}$, which is defined as~\cite{puertas2003simulation}

\begin{equation}
\label{eq:Dem}
\Psi_{l}=\sum_{k=0}^{l^{3}}|\rho_{k}-\rho|^{2}\,,
\end{equation}

\noindent where $\rho_{l}$ is the density in the $l-$th cell and $\rho=6\phi/(\pi d^{3})$ is the average particle number density. In the case of a homogeneous state $\Psi_{l}\rightarrow0$, while it grows in the presence of a phase separation. We study the evolution of the demixing parameter for $l=10$, calculating $\Psi_{10}$ during the deformation of the system. In this way, the simulation box is divided into $10^{3}=1000$ boxes. We checked that the results do not qualitatively depend on the specific choice of $l$, and hence, we study the normalized quantity $\Psi=\Psi_{l}/l^{3}$.

\begin{figure}[t]
\includegraphics[width=\linewidth]{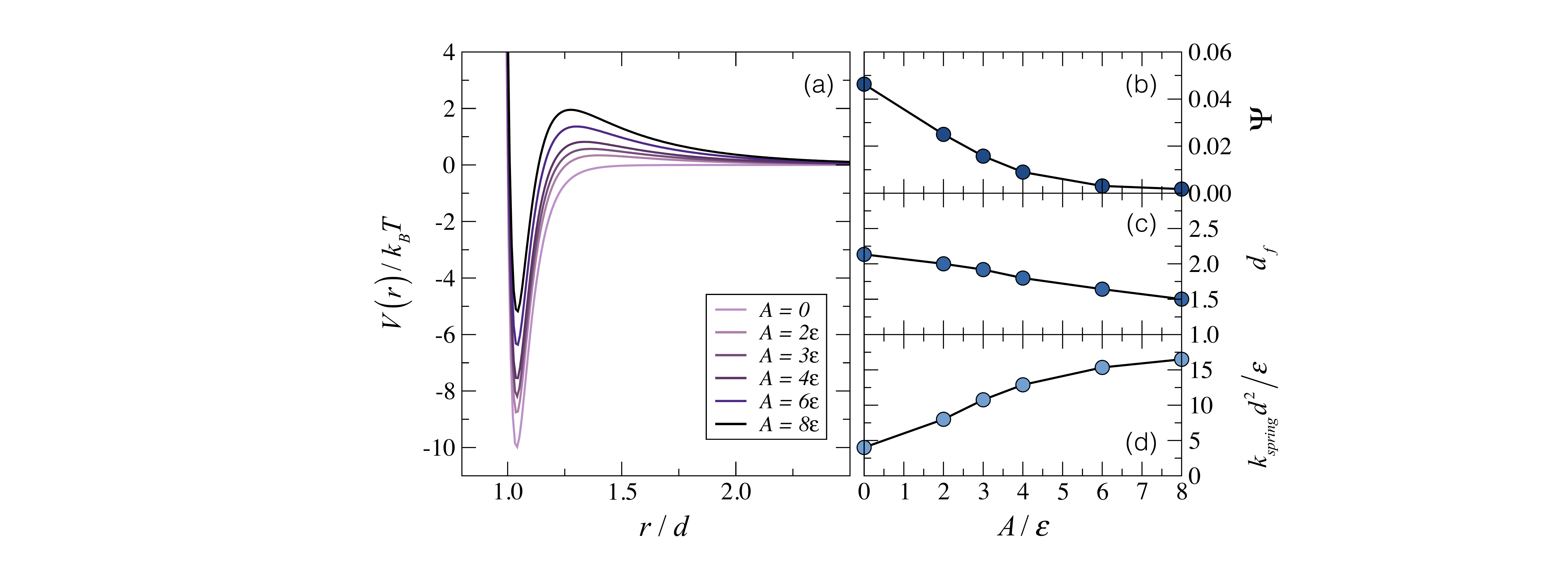}
\caption{(a) Total interaction potential $\beta V(r)$, (b) demixing parameter $\Psi$, (c) fractal dimension $d_{f}$, and (d) dimensionless effective spring constant $k_{spring} d^{2}/\epsilon$ of the gel for varying repulsive strength $A$, all in the absence of shear. }
\label{fig:Fig1}
\end{figure}

{\it Fractal dimension.} The study of the fractal dimension is performed using the box counting method~\cite{gagnepain1986fractal,griffiths2017local}. Thus, the simulation volume is divided into boxes of mesh size $r$ and we evaluate the number of cells $N(1/r)$ filled by the gel structure as a function of $1/r$. This is repeated for a series of mesh sizes, and hence, the fractal dimension is obtained from the slope of

\begin{equation} 
\label{eq:box_counting}
d_{f}=\frac{logN(1/r)}{log 1/r}\,.
\end{equation}

\noindent To obtain an accurate $d_{f}$, we identify in both, equilibrium and steady shear flow, the largest cluster (typically $N>40000$) whereas the smaller ones are discarded, and we compute its $d_{f}$. 

\section{III. RESULTS}
\subsection{A. Mechanical equilibrium}

We start by assessing the role of the repulsion strength on the behavior of the system in mechanical equilibrium, i.e., when $\dot{\gamma}=0$. As expected, the tendency of the system to phase separate is greatly reduced by increasing $A$, as shown in Fig.~\ref{fig:Fig1}(b): for $A=8\epsilon$ structural inhomogeneities are almost absent and we refer to this state as an equilibrium gel, whereas we call spinodal gel the one that is found in the absence of repulsion ($A=0$). The proximity to phase separation is reflected in the behavior of the gel fractal dimension $d_{f}$ (see Fig.~\ref{fig:Fig1}(c)): increasing repulsion, we promote the formation of strands~\cite{campbell2005dynamical,sciortino2005onedimensional}, thus reducing $d_{f}$ down to $\approx 1.5$. From the mechanical point of view, since we control the microscopic structure during the formation of the gel, we can also modify the stiffness of the bonds. This can be observed by defining an effective spring constant~\cite{dinsmore2006microscopic,zaccone2009elasticity,rocklin2018elasticity}

\begin{equation} 
\label{eq:kspring}
k_{spring} = \frac{k_{B}T}{\left\langle r^{2}\right\rangle -\left\langle r\right\rangle^{2}}\,,
\end{equation}

where $r$ is the distance between two bonded particles, i.e., they satisfy the constraint $r\leq r_{bond}$.  $r_{bond}$ is the bond distance defined by the position of the local maximum of $V(r)$, except for $A=0$ where we consider  $r_{bond}=1.5d$. $\left\langle r ^2\right\rangle$ and $\left\langle r \right\rangle^2$ are evaluated over all possible bonded pairs. The resulting effective spring constant $k_{spring}$ is shown in Fig.~\ref{fig:Fig1}(d), indicating that the network is stiffer in the presence of stronger repulsion.

\subsection{B. Shear response}

\begin{figure*}[t]
\includegraphics[width=0.8\linewidth]{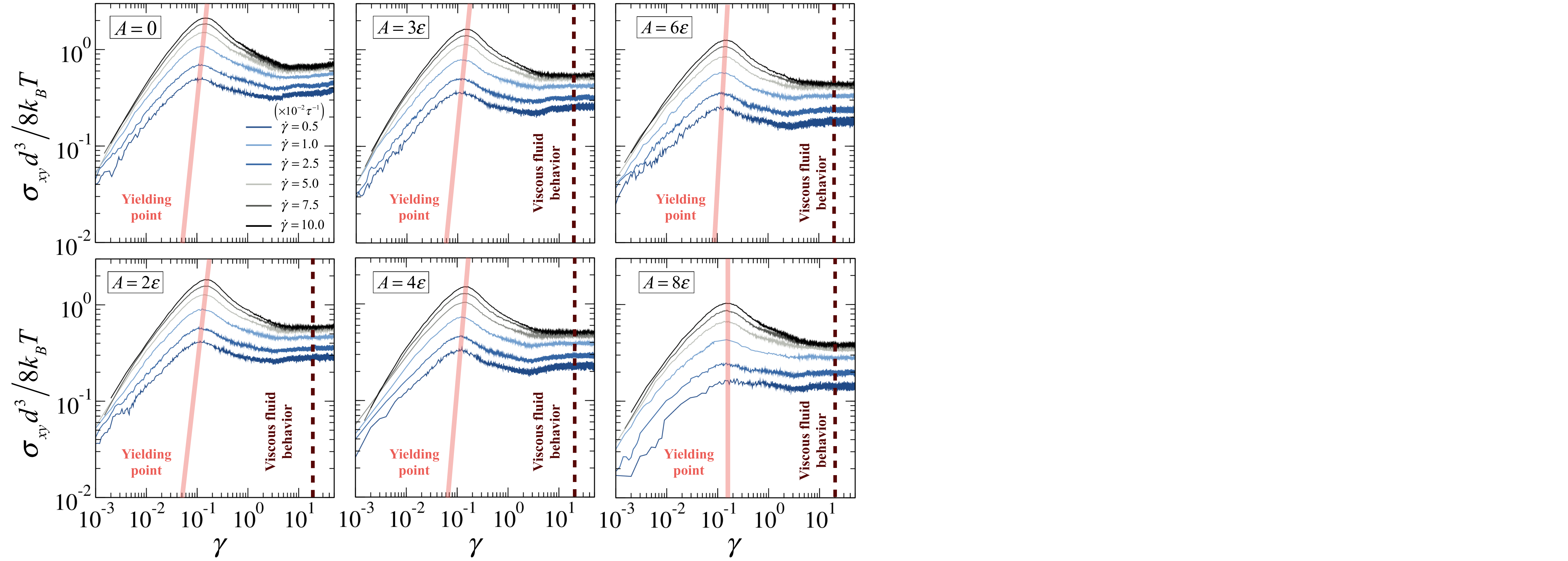}
\caption{Shear stress $\sigma_{xy}$ versus strain $\gamma$ for different shear rates $\dot{\gamma}$ and repulsive strengths $A$.  Yielding points $\gamma_{yield}$ are highlighted by solid lines, whereas the onset of steady-state viscous fluid behavior is indicated by dashed lines.} 
\label{fig:Fig2}
\end{figure*}

Next we apply a steady shear protocol and monitor the evolution of $\sigma_{xy}$ as a function of  strain $\gamma=\dot{\gamma}t$ for $\dot{\gamma}\in\left[10^{-5}-0.15\right]$, as reported in Fig.~\ref{fig:Fig2}. 
Independently of the value of $A$, we find the typical behavior of a viscoelastic material, being characterized by the presence of a yielding transition where $\sigma_{overshoot}$ and $\gamma_{yield}$ related to the maximum stored energy in the system and to the maximum deformation, respectively. 
The variation in the repulsion contribution induces important changes: while for a spinodal gel, the position of $\gamma_{yield}$ shows a dependence on $\dot{\gamma}$~\cite{johnson2018yield}, for an equilibrium gel such a dependence is essentially absent~\cite{ruiz2019rheological}. 
This can be attributed to the different organization of particles within the network: while for large values of $A$ bonded particles are more constrained to maintain a preferred distance, for low $A$ they can sufficiently reorganize themselves thus delaying the gel fracture to larger values of strain.

\subsection{C. Stress overshoot}

\begin{figure}[b]
\includegraphics[width=\linewidth]{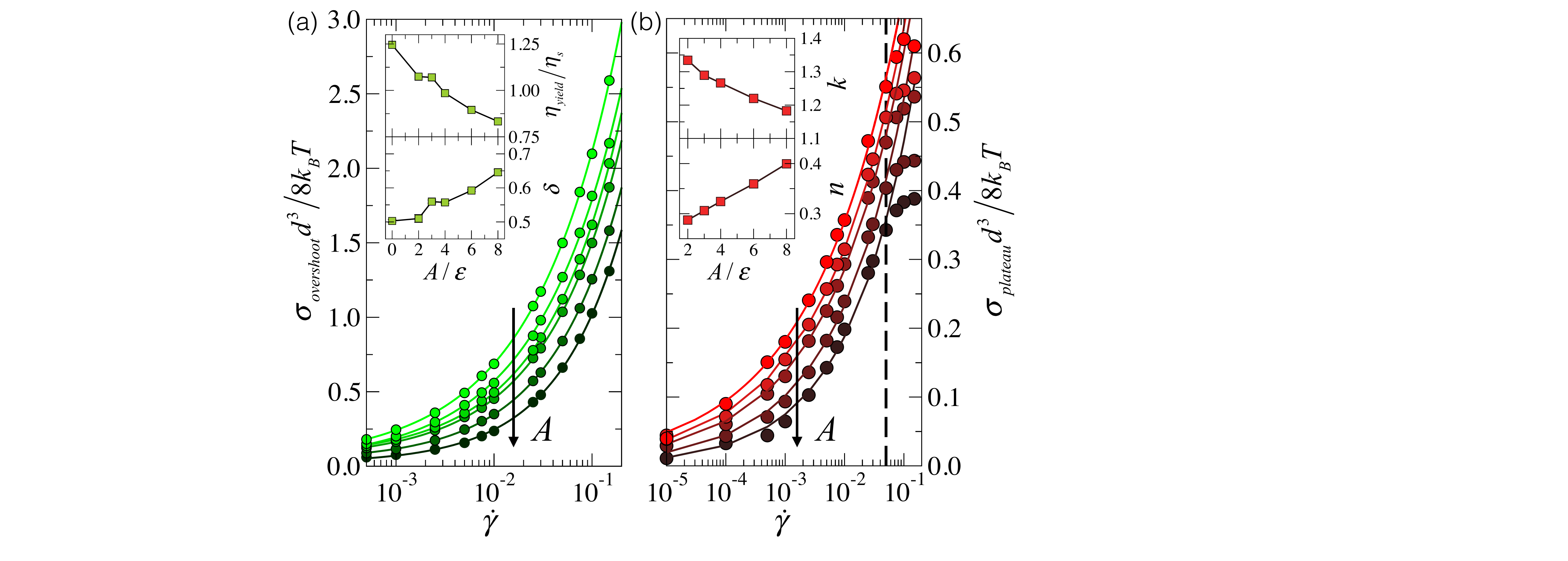}
\caption{(a) Stress overshoot $\sigma_{overshoot}$ and (b) stress plateau  $\sigma_{plateau}$ as a function of $\dot{\gamma}$ for different $A$ values. Symbols are numerical results, solid lines are fits with the power-law $\sigma_{overshoot}\sim \eta_{yield}\dot{\gamma}^{\delta}$ in (a) and with the Herschel-Bulkley model in (b). In the second case, the vertical dashed line signals the upper boundary of the fits. {\it Insets}: $A$-dependence of effective viscosity $\eta_{yield}$ (a-top), exponent $\delta$ (a-bottom), consistency index $k$ (b-top) and flow index $n$ (b-bottom).}
\label{fig:Fig3}
\end{figure}

Recently, the yielding transition of  amorphous materials has attracted a large interest ~\cite{wisitsorasak2012strength,urbani2017shear,leishangthem2017yielding,biroli2018liu,ozawa2018random}. Specifically regarding spinodal gels, different dependencies on $\phi$, on the interparticle attractive energy or on $\dot{\gamma}$ have been reported~\cite{groot1995dynamic,trappe2001jamming,koumakis2011two,ruiz2019rheological}. 
In addition, data were found to follow the power law $\sigma_{overshoot}\sim \eta_{yield}\dot\gamma^{\delta}$, where $\eta_{yield}$ can be interpreted as an effective viscosity of the system  when the structure begins to break~\cite{johnson2018yield}. This description is found to hold also for the present data, as shown in Fig.~\ref{fig:Fig3}(a), where the evolution of $\sigma_{overshoot}$ as a function of $\dot{\gamma}$ is reported for different values of $A$. By increasing the strength of repulsion, the energy accumulated within the structure, for the same variation of $\dot{\gamma}$, is significantly reduced. This is accompanied by an increase in the exponent $\delta$, indicating an enhanced bond stiffness (see inset of Fig.~\ref{fig:Fig3}(a)) similarly to what observed earlier for $k_{spring}$ (Fig.~\ref{fig:Fig1}(d)). 
Thus, the increase of the effective charges on the colloids modifies the response of the material from more ductile to more brittle. 
In addition, $\eta_{yield}$ is found to decrease with increasing $A$ (see inset of Fig.~\ref{fig:Fig3}(a)). This is a counterintuitive result: upon increasing the electrostatic repulsion, the gels, despite being stiffer, become less viscous. This happens because, for the present system, lowering the repulsion has the effect to enhance the tendency to phase separate and, thus, the gels become locally more compact, thereby displaying a  larger effective viscosity. 

\subsection{D. Viscous fluid behavior}

Once $\sigma_{overshoot}$ is overcome, $\sigma_{xy}$ decreases to a steady value indicating that the system exhibits viscous fluid properties. In this regime, $\sigma_{xy}$  reaches a long-time plateau (for large enough values of $A$) and is well described by the Herschel-Bulkley (HB) model~\cite{osswald2015polymer}, namely $\sigma_{plateau}\sim\sigma_{o}+k\dot{\gamma}^{n}$. Here $\sigma_{o}$ is the yield stress and $n$ is the flow index that characterizes the type of viscous behavior. While for Newtonian fluids  $n=1$, shear-thinning and shear-thickening behaviors are characterized by $n<1$ and $n>1$, respectively.  Finally $k$ is the consistency index, which plays a similar role to the viscosity for a non-Newtonian fluid. We report in  Fig.~\ref{fig:Fig3}(b) the evolution of $\sigma_{plateau}$ with $\dot{\gamma}$ for different values of $A$. Here, we exclude the case $A=0$ due to the lack of a clear long-time plateau (see Fig.~\ref{fig:Fig2}). In general, we find that all data are well-described by the HB model until $\dot{\gamma}\sim 0.5\tau^{-1}$, above which deviations are identified due to flow-induced inhomogeneities.
From the HB fits, we estimate $k$ and $n$, which are both found to linearly depend on $A$. In analogy to the features observed at the yielding point, we find that $n$ increases, while $k$ decreases. This brings to the remarkable finding that a stronger repulsion in the system increases its tendency to flow, even in steady state.

\begin{figure}
\includegraphics[width=0.905\linewidth]{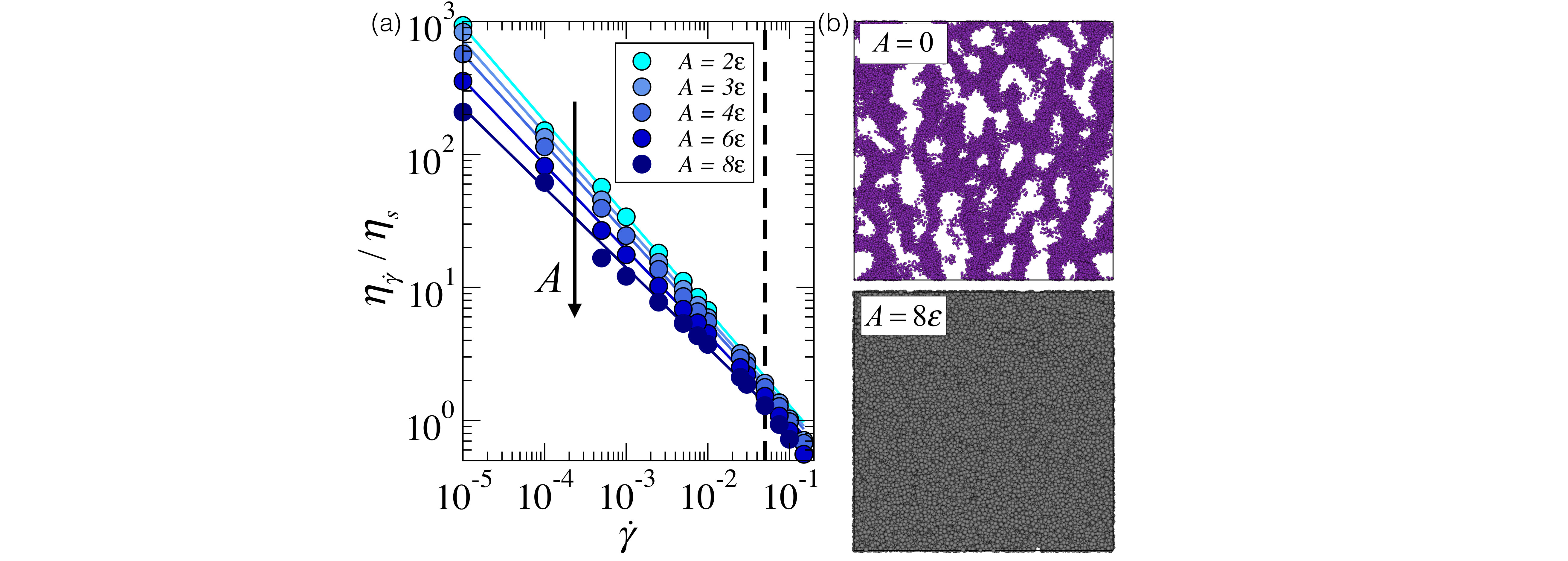}
\caption{(a) Steady-state viscosity $\eta_{\dot{\gamma}}$ as a function of $\dot{\gamma}$ for different repulsive strengths $A$. Symbols are numerical results, whereas solid lines are obtained by the HB fits of  $\sigma_{plateau}$ in Fig.~\protect\ref{fig:Fig3}(b). Dashed line indicates the upper boundary of the fits; (b) snapshots of the gel at $A=0$ and $A=8\epsilon$ in the presence of a shear flow with $\dot{\gamma}=10^{-2}\tau^{-1}$ and $\gamma=50$. The particle size was arbitrarily reduced for a better visualization.}
\label{fig:Fig4}
\end{figure}

The ability to flow under shear by increasing the electrostatic contribution is also  captured by the behavior of the viscosity, which can be directly calculated in the steady state as $\eta_{\dot{\gamma}}=\sigma_{plateau}/\dot{\gamma}$, reported in Fig.~\ref{fig:Fig4}(a) as a function of $\dot{\gamma}$. We find that, for the same value of $\dot{\gamma}$, the viscosity decreases (by almost a decade for small shear rates) with increasing  $A$, at odds with common expectations according to which a larger constraint to the system given by repulsive interactions should act against particle motion. This peculiar behavior originates from the presence of competing interactions in our system and, specifically, to the fact that a decrease in the repulsion does actually lead to an increase the overall effect of the short-range attraction. This can be visualized in the snapshots reported in Fig.~\ref{fig:Fig4}(b), corresponding to the gels at $A=0$ and $A=8\epsilon$ in the presence of shear with  $\dot{\gamma}=10^{-2}\tau^{-1}$ and $\gamma=50$. It is evident that the gels become more and more locally compact, but largely heterogeneous, as $A$ decreases.

\begin{figure}
\includegraphics[width=\linewidth]{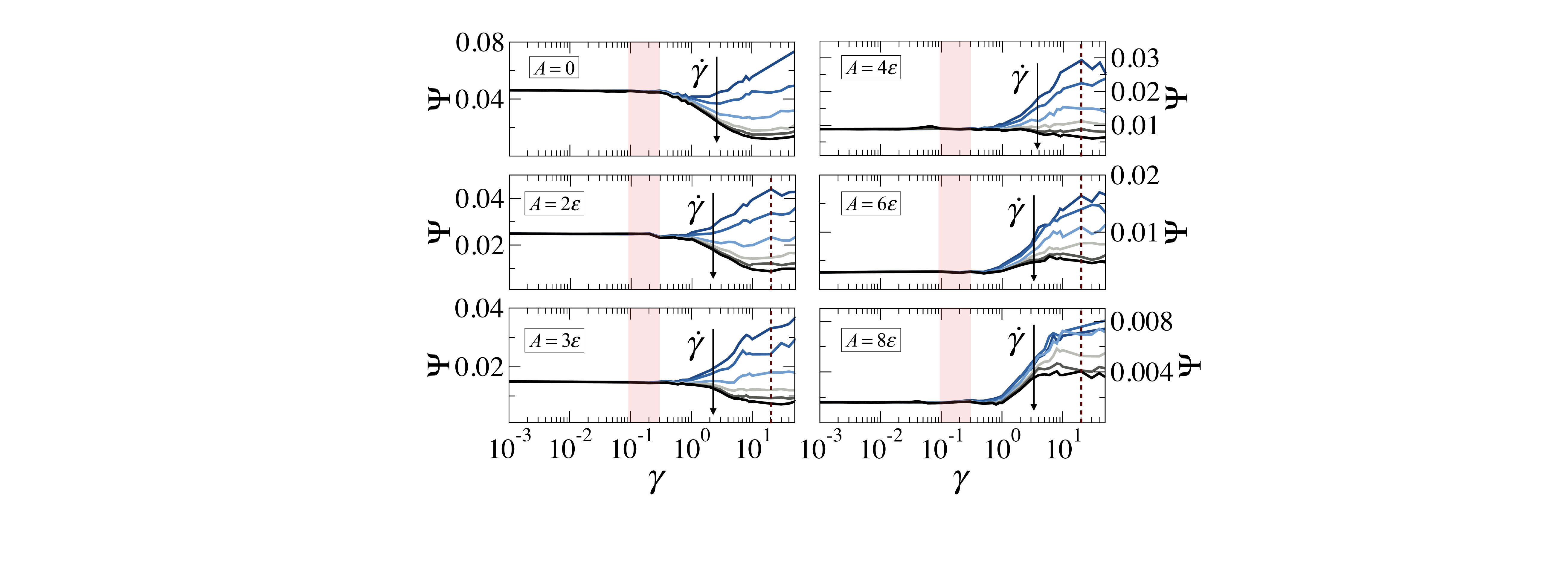}
\caption{Demixing parameter $\Psi$ versus strain $\gamma$ for different values of $\dot\gamma$ and $A$. Shaded areas indicate the position of $\gamma_{yield}$, while the vertical dashed lines mark the onset of the viscous fluid behavior. The color scheme is the same as in Fig.~\ref{fig:Fig2}.}
\label{fig:Fig5}
\end{figure}

To shed light on these fascinating findings, we follow the behavior of the demixing parameter $\Psi$ as a function of strain for different values of $A$ and $\dot{\gamma}$. As shown in Fig.~\ref{fig:Fig5}, $\Psi$ remains roughly constant for all studied systems until the yielding point, after which it begins to vary, reaching a large strain plateau when the system approaches a viscous steady state. We see that in the absence of long-range repulsion and up to $A\lesssim 3\epsilon$, the shear has the clear effect of reducing the tendency of the system to phase separate. Thus it reaches a viscous-like behavior, particularly for large shear rates, due to the breaking of the system into smaller clusters. On the other hand, the situation is reversed for $A\geq6\epsilon$ after which the system is actually pushed closer to phase separation independently of shear rate. This happens because the shear is able to effectively screen the repulsive interactions, giving the possibility to the particles to further accommodate attractive bonds, thus growing the aggregates that are present in the system with respect to the corresponding equilibrium state. For intermediate values of $A$, such tendency to phase separate can be further tuned, finding, for example, situations where no changes with respect to the equilibrium state are found (see for example $A=4\epsilon$ and large shear rates). 

\subsection{E. Fractal dimension}

\begin{figure}
\includegraphics[width=\linewidth]{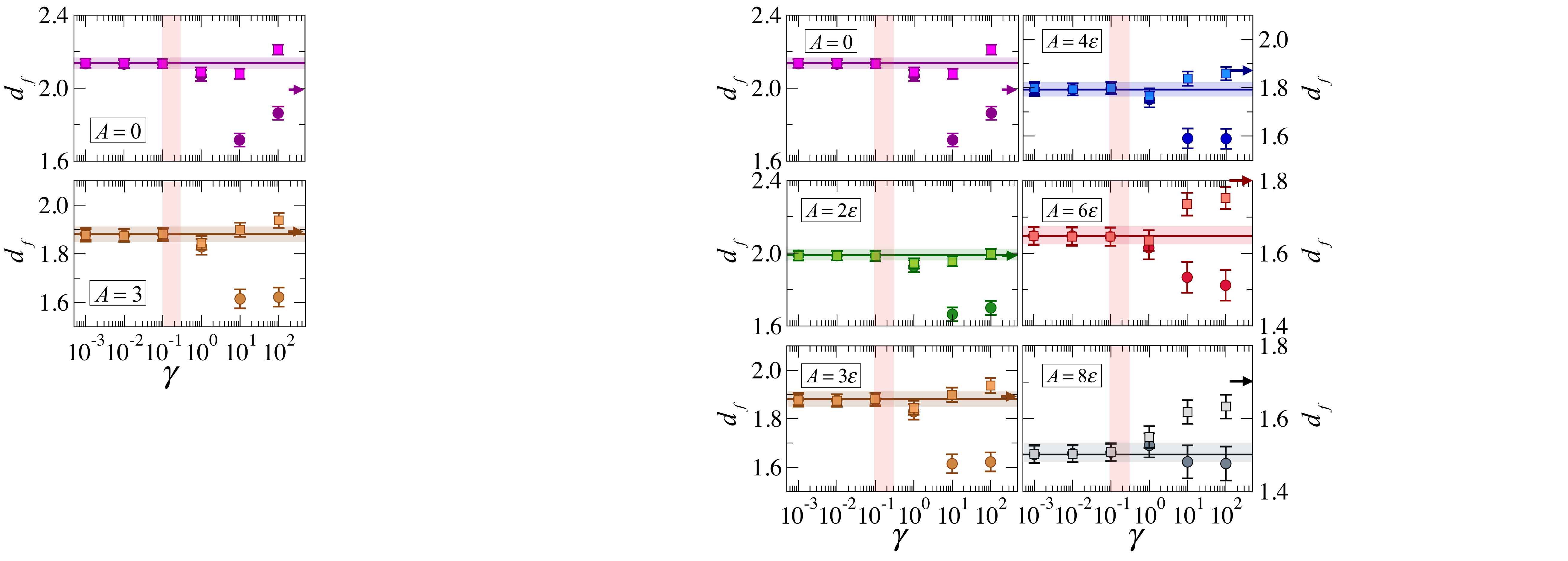}
\caption{Fractal dimension $d_{f}$ as a function of strain $\gamma$ for $\dot{\gamma}=10^{-2}\tau^{-1}$ (squares) and $\dot{\gamma}=10^{-1}\tau^{-1}$ (circles) and different values of $A$. Horizontal lines/shaded areas indicate the value of $d_{f}$ and the associated error in equilibrium; vertical shaded areas highlight the position of $\gamma_{yield}$. Arrows point to the value of $d_{f}$ obtained from the $\delta$ exponent  (see text).}
\label{fig:Fig6}
\end{figure}

Another intriguing result is given by the analysis of the fractal dimension $d_f$ of the gels, that is reported as function of strain in Fig.~\ref{fig:Fig6} for different $A$ and for two representative values of the shear rate. As expected, for $\gamma<\gamma_{yield}$, $d_{f}$ coincides within the error bars with the corresponding value in mechanical equilibrium. Only for $\gamma\geq\gamma_{yield}$ there is an effect of the shear flow on $d_{f}$. Strikingly we find that this can either grow or decrease upon changing shear rate. Indeed, for small values of $\dot{\gamma}$, the fractal dimension becomes larger with respect to the equilibrium value, while for large $\dot{\gamma}$ it decreases. The two scenarios can be explained by the subtle balance between shear, attractive, and repulsive interactions. 
Indeed, for small shear rates, the external flow is able to counterbalance the repulsion at large distances, without considerably affecting the attractive bonds between the colloids. This enhances the proximity to the phase separation, as also seen from the behavior of the demixing parameter. On the other hand, for sufficiently high shear rates, the flow becomes too strong, melting the gel and effectively breaking attractive bonds. Under these conditions, layering effects become dominant, thus lowering the fractal dimension. This is confirmed by the fact that the lowest values of $d_f$ $\sim 1.5/1.6$ at high shear rates are almost identical for all studied values of $A$. Instead the low-shear-rate value of $d_f$  is found to clearly depend on the underlying potential.  It was recently proposed that in polymer gels the fractal dimension is connected to the exponent $\delta$ describing the power-law dependence of the stress overshoot at yielding, discussed in Fig.~\ref{fig:Fig3}(a), via the relationship  $d_f=3-2\delta$~\cite{groot1995dynamic,park2013structural}. We are now in the position to verify this hypothesis for colloidal gels, comparing  $d_f$, directly calculated from simulations, with such a prediction, also reported in Fig.~\ref{fig:Fig6}. Interestingly, we find that the last agrees reasonably well with the asymptotic value of $d_f$ for large strains at all $A$. Thus, we conclude that the relation between $d_f$ and $\delta$ mentioned above seems to apply also to colloidal gels but only for small-enough shear rates and for very large strains, where it can be used as a guide to predict the final value of $d_f$. These results clearly indicate that the repulsion strength can be used as a novel control parameter to tune the dynamical and structural properties of colloidal gels, controlling the resulting viscosity of the system and being able to drive aggregation of the particles into structures with a wide range of fractal dimensions.  

\section{IV. CONCLUSIONS}

In summary, we present extensive numerical results of colloidal gels under start-up shear upon changing the contribution of the long-range electrostatic repulsion in the underlying interaction potential. We find that increasing the strength of the repulsion, despite stiffening the network, actually lowers its viscosity. This is entirely attributable to the presence of competing interactions and can be rationalized as a shear-homogenization effect. Indeed, the external field is able to push the system closer to macroscopic phase separation that is avoided in equilibrium by the additional repulsive contribution. We also show that simple changes in the electrostatic potential directly manifest in the rheological and structural properties of the gels, giving rise to networks with very different fractal dimensions. These can be made either larger or smaller with respect to the system in mechanical equilibrium. It will be extremely interesting to verify whether such a manipulation of the gel properties can be obtained in the laboratory, by appropriately changing the effective colloidal charge, to study its influence on the rheological properties both from a fundamental point of view and for a wide range of industrial and technological problems.

We acknowledge support from the European Research Council (ERC Consolidator Grant 681597, MIMIC) 
and from Sapienza University of Rome through the SAPIExcellence program.
%


\end{document}